\documentclass[journal]{IEEEtran}
\usepackage{epstopdf}
\usepackage{multirow}
\usepackage{cite}
\usepackage{stfloats}
\usepackage{color}
\usepackage{epsfig}
\usepackage{graphicx}
\usepackage{psfig}
\usepackage{epsf}
\usepackage{slashbox}
\usepackage{float}
\usepackage{amssymb }
\usepackage[cmex10]{amsmath}
\usepackage{algorithm} 
\usepackage{algorithmic}
\usepackage{setspace}
\usepackage{amsfonts}
\usepackage{amsmath}
\usepackage{booktabs}
\usepackage{amsmath,bm}
\usepackage[justification=centering]{caption}
\title{Power Beacon Energy Consumption Minimization in Wireless Powered Backscatter Communication Networks}
\author{Haohang Yang, Yinghui Ye, Kai Liang, Xiaoli Chu,~\IEEEmembership{Senior Member,~IEEE}
\thanks{Haohang Yang and Xiaoli Chu are with the Department of Electronic and Electrical Engineering, University of Sheffield, Sheffield, S1 4ET, United Kingdom (e-mail: hyang42@sheffield.ac.uk, x.chu@sheffield.ac.uk).}
\thanks{Yinghui Ye is with the Shaanxi Key Laboratory
of Information Communication Network and Security, Xi'an University
of Posts \& Telecommunications, China (e-mail: connectyyh@126.com).}
\thanks{Kai Liang is with the State Key Laboratory of Integrated Service Networks, Xidian University, Xi’an 710071, China (e-mail: kliang@xidian.edu.cn).}
}
\begin{document}
\maketitle
\begin{abstract}
Internet-of-Things (IoT) networks are expected to support the wireless connection of massive energy limited IoT nodes. The emerging wireless powered backscatter communications (WPBC) enable IoT nodes to harvest energy from the incident radio frequency signals transmitted by a power beacon (PB) to support their circuit operation, but the energy consumption of the PB (a potentially high cost borne by the network operator) has not been sufficiently studied for WPBC. In this paper, we aim to minimize the energy consumption of the PB while satisfying the throughput requirement per IoT node by jointly optimizing the time division multiple access (TDMA) time slot duration and backscatter reflection coefficient of each IoT node and the PB transmit power per time slot. As the formulated joint optimization problem is non-convex, we transform it into a convex problem by using auxiliary variables, then employ the Lagrange dual method to obtain the optimal solutions. To reduce the implementation complexity required for adjusting the PB’s transmit power every time slot, we keep the PB transmit power constant in each time block and solve the corresponding PB energy consumption minimization problem by using auxiliary variables, the block coordinated decent method and the successive convex approximation technique. Based on the above solutions, two iterative algorithms are proposed for the dynamic PB transmit power scheme and the static PB transmit power scheme. The simulation results show that the dynamic PB transmit power scheme and the static PB transmit power scheme both achieve a lower PB energy consumption than the benchmark schemes, and the former achieves the lowest PB energy consumption.
\end{abstract}


\section{INTRODUCTION}
\IEEEPARstart{W}{ith} the fast development and wide adoption of the Internet-of-Things (IoT) systems, a great number of low-power IoT nodes are being deployed to collect data and deliver them to the information fusion (IF), e.g, a gateway, for information processing \cite{9151232}. However, due to the	cost and form factor constraints, mobile IoT devices are typically powered by batteries with a limited capacity, which can be quickly drained by power-hungry radio frequency (RF) components, e.g., oscillators \cite{7397856}. To overcome the battery imposed limit, wireless powered backscatter communications (WPBC) \cite{9151232,8253544} have been proposed to allow an IoT node to modulate and backscatter the incident RF signals transmitted by a power beacon (PB) to carry the IoT node's information to the destination node, while harvesting energy from the PB's RF signals to support the circuit operation.

Various resource allocation schemes for WPBC have been developed to optimize different performance metrics, e.g., sum throughput, user-centric and system-centric energy efficiency (EE) \footnote{User-centric EE refers to the ratio of the throughput to the energy consumption cost by a user, while the system-centric EE is the EE of the whole network, which includes all the users, PBs and other communication equipments.} under the energy-causality constraint per IoT node. However, there is no study working on the energy efficient resource allocation scheme for WPBC from the network operator's perspective, e.g., minimizing the energy consumption of the PB. Such a scheme is vital for the deployment of WPBC in IoT networks, this is because the energy consumption of the user side is usually low and can be satisfied by the harvested energy in WPBC, but the energy consumption of the PB has significant impacts on the cost of the network operator. Motivated by this observation, in this paper, we design an energy efficient resource allocation scheme that minimizes the energy consumption of the PB while satisfying the throughput requirement per IoT node in a WPBC network with multiple IoT nodes.

\subsection{Related Works and Motivations}
Some related works on BackCom are listed as follows. In \cite{7876867}, using tools from the stochastic geometry and modeling the IoT nodes and the PB as the random Poisson cluster process, respectively, the authors derived the successful transmission probability of a large-scale WPBC network. It revealed that the average network capacity (or the successful transmission rate of a backscatter link) first increases and then decreases with the increasing duty cycle (or the backscatter reflection coefficient). The authors of \cite{8093703} maximized the throughput of an IoT node for single-user, single-carrier WPBC by jointly optimizing the backscatter reflection coefficient and the time for energy harvesting and that for backscatter communication (BackCom). Subsequently, this work was extended to a single-user, multi-carrier WPBC network, where the total throughput was maximized by jointly optimizing the transmit power on each subcarrier, the backscatter reflection coefficient, the energy harvesting (EH) time and the BackCom time \cite{9052662}. The PB’s signal waveform design and the associated tradeoff between the harvested energy of the IoT node and the signal-to-interference-plus-noise ratio (SINR) of the IoT node to the associated receiver link were studied for a single-user WPBC network in \cite{7950915} and for a multi-user WPBC network in \cite{8527670}. In \cite{9319216}, the rate-energy tradeoff was studied for both power splitting and time switching-based WPBC networks. The authors of \cite{8730429,8947165} developed resource allocation schemes to maximize the system-centric energy efficiency, which was defined as the ratio of the total throughput of the network to the total energy consumption of a network, in a WPBC system with or without quality of service (QoS) constraints of IoT nodes, respectively. In \cite{8957355}, the authors jointly optimized the backscatter reflection coefficient of each IoT node and the transmit power of a PB to maximize the minimum energy efficiency of all the IoT nodes, subject to the QoS requirement per IoT node.

The combination of WPBC and active transmissions (AT), where the IoT nodes convey information via hybrid BackCom and AT using the harvested energy, has also been studied. In \cite{8802296}, the Stackelberg game was employed to jointly maximize the PB’s revenue and the IoT nodes’ utility function. In \cite{8340034}, the time allocated for EH, BackCom, and AT was optimized to maximize the total throughput of a  multi-user hybrid BackCom-AT network. The sum throughput of all IoT nodes was maximized by optimizing the time allocation for EH, BackCom and AT in hybrid WPBC-AT-based heterogeneous networks \cite{7981380} and hybrid WPBC-AT-based cognitive radio networks \cite{8327597}, respectively. To ensure the throughput fairness among IoT nodes in a hybrid BackCom-AT network, the authors in \cite{9161012} solved a max-min throughput problem by jointly optimizing the backscatter reflection coefficient, the time for EH, BackCom and AT, and the transmit power of AT for each IoT node. Hybrid WPBC-AT was also employed in relay networks to enhance the achievable throughput of the IoT nodes by jointly optimizing the AT time and transmit power of the relay node \cite{8943100} or by optimizing the backscatter reflection coefficient of the relay node \cite{8922800}. The authors of \cite{8672817} maximized the user-centric energy efficiency, which was defined as the ratio of the throughput of an IoT node to the energy consumption of the IoT node, in a single-user hybrid WPBC-AT-enabled cognitive radio network \cite{8672817}. In \cite{9159908}, the total user-centric energy efficiency of all IoT nodes was maximized by jointly optimizing the transmit power of the PB and each IoT node's time for BackCom and AT under a non-linear energy harvesting model.

We note that the existing works on energy efficient or spectral efficient resource allocation schemes for WPBC are mainly from the system perspective with the assumption of signal user \cite{8730429,8947165} or the user perspective \cite{8672817,9159908}. Also, the energy consumption of the operator, e.g., a PB, is much higher than that of the users in WPBC networks due to their passive backscatter circuit. Thus, the energy consumption minimization scheme from the network operator's perspective with multiple users has not been studied and this motivates this work.

\subsection{Contributions}
In this paper, we consider a WPBC network of multiple IoT nodes, where in each time block, the IoT nodes take turns to modulate their own information on the incident RF signal from the PB and backscatter it to the IF following a time-division multiple-access (TDMA) protocol, and devise resource allocation schemes to minimize the energy consumption of the PB while satisfying the throughput requirement of each IoT node. Our main contributions are summarized as follows.

\begin{itemize}
\item The energy consumption of the PB in each time block is minimized by jointly optimizing the PB transmit power per time slot, the TDMA time slot duration and backscatter reflection coefficient of each IoT node, while guaranteeing the throughput requirements of the IoT nodes. By solving this joint optimization problem, we propose a dynamic PB transmit power scheme, where the PB adjusts its transmission power in each TDMA time slot according to the channel condition from the corresponding IoT node to the IF and to itself to minimize the PB's energy consumption in a time block. Considering the implementation complexity required for adjusting the PB's transmit power every time slot, we also propose a reduced-hardware complexity static PB transmit power scheme, where the PB transmit power remains constant in a transmission block and is optimized to minimize the PB energy consumption in each time block.

\item Due to multiple coupled variables in the objective function and the constraints, the two formulated optimization problems are non-convex and hard to solve directly. In the dynamic PB transmit power scheme, we use two auxiliary variables to transform the original optimization problem into a convex form and prove the convexity. Although the problem is transformed to be convex, we cannot simply use CVX tool to solve it due to the fractional form in some constraints. Instead, we employ the Lagrange dual method to solve it and derive the closed-form expressions of the optimal PB transmit power per time slot and the optimal backscatter reflection coefficients of the IoT nodes. In the static PB transmit power scheme, since the non-convex optimization problem cannot be transformed into a convex form, we employ the successive convex approximation (SCA) technique and block coordinated decent (BCD) method to solve the problem and obtain sub-optimal solutions of the PB transmit power and the IoT nodes' time slot durations and backscatter reflection coefficients. Based on the above solutions, we propose two iterative algorithms for the dynamic PB transmit power scheme and the static PB transmit power scheme, respectively.

\item We obtain and theoretically prove the following key insight. In the dynamic PB transmit power scheme, the PB energy consumption is minimized when the entire time block is used up for EH and BackCom by all the IoT nodes, but it may not be the case for the static PB transmit power scheme. The convergence and the run time of the proposed algorithms are analysed. Simulation results are presented to evaluate the convergence and the PB energy consumption performance of the proposed algorithms in comparison with the benchmark schemes that maximize the EE or the throughput.
\end{itemize}

The rest of the paper is organized as follows. The system model is presented in Section II. In Section III and Section IV, the PB energy consumption minimization problems for the two schemes are formulated and solved, respectively. Section V analyses the convergence and the run time of the proposed algorithms. In Section VI, the numerical results are presented. Section VII concludes the paper.

\begin{figure}
\centerline{\includegraphics[width=3.5in]{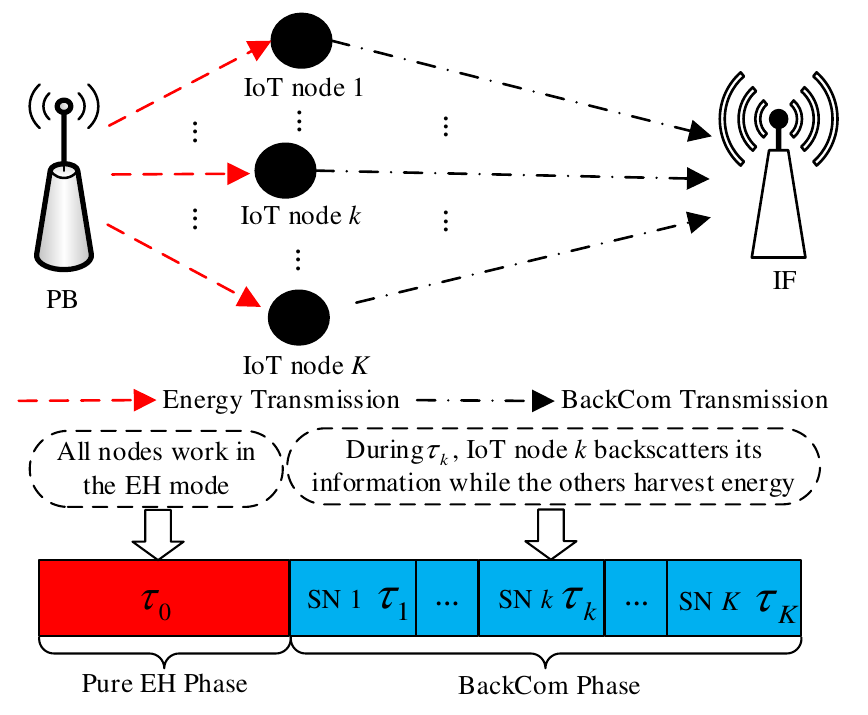}}
\caption{System model and time structure.}
\end{figure}

\section{System Model}
As shown in Fig. 1, we consider a WPBC network, which consists of one PB, $K$ IoT nodes and one IF. Both the PB and IF have stable power supply, while the $K$ IoT nodes are equipped with backscatter circuits and they are battery powered and energy-constrained. To minimize the battery power consumption at the IoT nodes, each IoT node harvests energy from the incident signal transmitted by the PB to support its own circuit operation and conveys  IoT nodes' information to the IF via BackCom following a TDMA protocol. We assume that all devices are equipped with a single antenna. We consider a block flat-fading channel model, where all the channel gains keep constant in one transmission block, but may change across adjacent transmission blocks. Each transmission block has a duration of $T$ seconds, which is shorter than the channel coherence time. As channel estimation is outside the scope of this paper, we assume that the perfect channel state information (CSI) of links from the IoT nodes to the PB and the IF can be obtained by the PB via an advanced channel estimation scheme at the beginning of each transmission block \cite{8320359,8746230,8618337}.

There are two phases within each transmission block\footnote{Although BackCom allows the IoT nodes realizing the information transmission and EH simultaneously, the harvested energy may not enough support the circuit operation due to the low EH efficiency. Thus, we schedule one phase for pure EH. }, i.e., the energy harvesting phase (with a duration of $\tau_0$) and the BackCom phase. In order to reduce the co-channel interference and the complexity of the IF, the BackCom phase is further divided into $K$ slots $\left\{ {{\tau _k},k = 1,2, \ldots ,K} \right\}$ and $\tau_k$ is allocated to the $k$-th IoT node for performing BackCom.

In the EH  phase $\tau_0$, the PB broadcasts energy signals, and the received signal at the $k$-th IoT node is expressed as
\begin{align}\label{a1}
{y_k} = \sqrt {{P_{0}}{h_k}} {x_s} + {w_{k,p}}\approx \sqrt {{P_{0}}{h_k}} {x_s},
\end{align}
where $x_s$ is the energy signal following a standard circularly symmetric complex Gaussian distribution, i.e., ${x_s} \sim \mathcal{CN}\left( {0,1} \right)$, ${P_{0}}$, $h_k$, and ${w_{k,p}}$ are the transmit power of the PB within $\tau_0$, the channel gain from  the PB to the $k$-th IoT node, and the thermal noise of the $k$-th IoT node, respectively. In \eqref{a1}, the approximation is obtained due to the fact that the BackCom circuit only consists of passive components and the value of $w_{k,p}$ is negligible.

In $\tau_0$, all IoT nodes harvest energy from the broadcasted energy signal. For a practical energy harvester, the harvested power is a non-linear function of the input power. Considering a  non-linear EH model \cite{7999248}, the harvested energy of the $k$-th IoT node within $\tau_0$ can be calculated as
\begin{align}\label{1}
E^k_{\tau_0}=\left( {\frac{{a{P_0}{h_k} + d}}{{{P_0}{h_k} + v}} - \frac{d}{v}} \right){\tau _0},
\end{align}
where  $a$, $d$ and $v$ are the parameters of the non-linear EH model and can be determined by  the experimental fitting approach \cite{7999248}, and ${P_0}{h_k}$ is the input power. 

In the BackCom phase, the PB continues to broadcast energy signals and each IoT node takes turns (i.e., in their allocated time slot) to work in the BackCom mode, while all the other IoT nodes continue to harvest energy. In time slot $\tau_k$, the harvested energy of the $i$-th IoT node, $i = \left\{ {1,2, \ldots ,K,i \ne k} \right\}$, is expressed as
 \begin{align}\label{2}
 E^i_{\tau_k}=\left( {\frac{{a{P_k}{h_i} + d}}{{{P_k}{h_i} + v}} - \frac{d}{v}} \right){\tau _k},
 \end{align}
 where ${P_k}$ is the transmit power of the PB within $\tau_k$.

At the $k$-th IoT node in time slot $\tau_k$, the incident signal from the PB is split into two parts via a power reflection coefficient $\beta_k$, i.e., the part  $\sqrt {{\beta _k}{P_0}{h_k}} {x_s}$ for carrying the $k$-th IoT node's information to the IF, and the remaining part $\sqrt {\left( {1 - {\beta _k}} \right){P_0}{h_k}} {x_s}$ flowing into the EH circuit, where $0\le \beta_k\le 1, \forall k$. Then the achievable throughput of the backscatter link from the $k$-th IoT node to the IF and the harvested energy by the $k$-th IoT node within $\tau_k$ can be written as, respectively,
 \begin{align}\label{3}
 R_k^{{\tau _k}} = W{\tau _k}{\log _2}\left( {1 + \frac{{\xi\beta_k {P_k}{h_k}{g_k}}}{{W{N_0}}}} \right),
 \end{align}
 \begin{align}\label{4}
  E^k_{\tau_k}=\left( {\frac{{a\left( {1 - {\beta _k}} \right){P_k}{h_k} + d}}{{\left( {1 - {\beta _k}} \right){P_k}{h_k} + v}} - \frac{d}{v}} \right){\tau _k},
 \end{align}
where ${g_k}$ is the channel gain between the $k$-th IoT node and the IF, $W$ is the channel bandwidth, $N_0$ is the  noise power spectrum density, and $0<\xi<1$ is a parameter to describe the performance gap between the BackCom and the Shannon capacity\footnote{Due to the passive modulation, the throughput of BackCom is much less than the Shannon capacity. We note that in some  works, the achievable rate of BackCom is approximated as a constant that is determined by the  experimental testing. Such an approach is valid when the transmit power of a PB and the reflection coefficient keep unchange. In this work, since we jointly optimize the transmit power of the PB and the reflection coefficient of per IoT node, we use the Shannon capacity to approximate BackCom capacity and  a parameter is adopted  to reflect their performance gap \cite{9161012}.}. Please note that the IF receives the  signals from the $k$-th IoT node and  the PB simultaneously, but the received  signal from the PB can be perfectly removed by using successive interference cancellation (SIC). This is because the PB does not transmit information and the energy signal $x_s$ can be predefined and known by the IF. In this case, once the channel coefficient between the PB and the IF is estimated, SIC can be performed.

Based on \eqref{1}, \eqref{2} and \eqref{4}, the total harvested energy of the $k$-th IoT node during one transmission block is given by
\begin{align}\notag \label{5}
E_k^{\rm{total}}&=\left( {\frac{{a\left( {1 - {\beta _k}} \right){P_k}{h_k} + d}}{{\left( {1 - {\beta _k}} \right){P_k}{h_k} + v}} - \frac{d}{v}} \right){\tau _k}\\
&+\sum\limits_{i = 0,i \ne k}^K {\left( {\frac{{a{P_i}{h_k} + d}}{{{P_i}{h_k} + v}} - \frac{d}{v}} \right){\tau _i}}.
\end{align}

\section{PB Energy Consumption Minimization with Dynamic Transmit Power}
In this section, we introduce a dynamic PB transmit power scheme, where the PB’s transmit power can be adjusted in every time slot, and formulate a joint optimization problem to minimize the energy consumed by the PB while meeting the throughput requirement of each IoT node, and transform it into a convex problem and solve it by a Lagrange dual method.

\subsection{Problem Formulation}
We propose to minimize the energy consumption of the PB by jointly optimizing the time slot durations for EH and BackCom, and the reflection coefficient of each IoT node, and the transmit power of the PB at each time slot, subject to the minimum throughput requirement per IoT node. Mathematically, the problem can be written as
\begin{subequations}
\begin{align}
&{{\cal P}_1}:\;\;\;\;\mathop {\min }\limits_{{\tau _0},{\bm{\tau}},{\bm{\beta }},{P_0},{\bm{P}}} \;\sum\limits_{i = 0}^K {{P_i}{\tau _i}} \\
{\rm{s}}{\rm{.t}}{\rm{.}}\;\;&{\rm{C}}1:\; W{\tau _k}{\log _2}\left( {1 + \frac{{\xi\beta_k {P_k}{h_k}{g_k}}}{{W{N_0}}}} \right) \ge R_k^{\min },\;\forall k,\\ \notag
&{\rm{C}}2:\left( {\frac{{a\left( {1 - {\beta _k}} \right){P_k}{h_k} + d}}{{\left( {1 - {\beta _k}} \right){P_k}{h_k} + v}} - \frac{d}{v}} \right){\tau _k}\\
&\;\;\;\;\;+\sum\limits_{i = 0,i \ne k}^K {\left( {\frac{{a{P_i}{h_k} + d}}{{{P_i}{h_k} + v}} - \frac{d}{v}} \right){\tau _i}} \ge {p_{c,k}}{\tau _k},\forall k,\\
&{\rm{C}}3:{\tau _0},\;{\tau _k} \ge 0,\forall k, \sum\limits_{k = 0}^K {{\tau _k}}  \le T,\\
&{\rm{C}}4:0 \le {\beta _k} \le 1,\\
&{\rm{C}}5:0 < {P_0},{P_k} \le {P_{\max }},\forall k,
\end{align}
\end{subequations}
where ${\bm{\tau }} = \left[ {{\tau _1},{\tau _2}, \cdots ,{\tau _K}} \right]$, ${\bm{\beta }} = \left[ {{\beta _1},{\beta _2}, \cdots ,{\beta _K}} \right]$, ${\bm{P }} = \left[ {{P _1},{P _2}, \cdots ,{P _K}} \right]$, and
$R_k^{\min }$ and ${p_{c,k}}$ are the minimum required throughput and the circuit power consumption of the $k$-th IoT node, respectively.

In  ${\mathcal{P}_1}$, constraint ${\rm{C}}1$ guarantees that the achievable throughput of the $k$-th IoT node is above $R_k^{\min }$. Constraint ${\rm{C}}2$ requires that the consumed energy by the $k$-th IoT node should not exceed its harvested energy.   ${\rm{C}}3$ constrains the EH time and the BackCom time. Constraints ${\rm{C}}4$ and  ${\rm{C}}5$ set the value range of the reflection coefficient of the $k$-th IoT node and the transmit power of the PB, respectively. ${\mathcal{P}_1}$ is non-convex due to the coupled variables in the objective function, and constraints ${\rm{C}}1$ and ${\rm{C}}2$. Moreover, ${\rm{C}}2$ contains nonlinear fractional functions due to the use of the non-linear EH model. Therefore, problem ${\mathcal{P}_1}$ is non-convex and cannot be directly solved by using existing convex tools, e.g, CVX. In what follows, we will transform ${\mathcal{P}_1}$ into a convex problem and propose an iterative algorithm to solve it.

\subsection{Problem Transformation}
To decouple the optimization variables $P_i$ and $\tau_i$, we construct a series of  auxiliary variables $\theta_i=P_i\tau_i$, $i=0,1,..., K$, and substitute ${P_i} = \frac{{{\theta _i}}}{{{\tau _i}}}$ into ${\mathcal{P}_0}$, which results in
\begin{subequations}
\begin{align}
&{{\cal P}_{1.1}}:\;\;\;\;\mathop {\min }\limits_{{\tau _0},{\bm{\tau}},{\bm{\beta }},{\theta_0},{\bm{\theta}}} \;\sum\limits_{i = 0}^K {\theta_i} \\
{\rm{s}}{\rm{.t}}{\rm{.}}\;\;&{\rm{C}}1': W{\tau _k}{\log _2}\left( {1 + \frac{{\xi {\beta _k}{\theta _k}{h_k}{g_k}}}{{{\tau _k}W{N_0}}}} \right) \ge R_k^{\min },\;\forall k,\\ \notag
&{\rm{C}}2':\left( {\frac{{a\left( {1 - {\beta _k}} \right)\frac{{{\theta _k}}}{{{\tau _k}}}{h_k} + d}}{{\left( {1 - {\beta _k}} \right)\frac{{{\theta _k}}}{{{\tau _k}}}{h_k} + v}} - \frac{d}{v}} \right){\tau _k}\\
&\;+\sum\limits_{i = 0,i \ne k}^K {\left( {\frac{{a\frac{{{\theta _i}}}{{{\tau _i}}}{h_k} + d}}{{{{\frac{{{\theta _i}}}{{{\tau _i}}}}}{h_k} + v}} - \frac{d}{v}} \right){\tau _i}}  \ge {p_{c,k}}{\tau _k},\forall k,\\
&{\rm{C}}5':0 < {\theta_0} \le P_{\rm{max}}{\tau _0},0 <{\theta_k} \le P_{\rm{max}}{\tau _k},\forall k,\\
&{\rm{C}}3,\;{\rm{C}}4,\notag
\end{align}
\end{subequations}
where ${\bm{\theta}} = \left[ {{\theta _1},{\theta _2}, \cdots ,{\theta _K}} \right]$.

In ${{\cal P}_{1.1}}$, the objective function (8a) is linear and thus is more tractable than (7a). However, the reflection coefficient $\beta_k$ is still coupled with $\theta_k$ in ${\rm{C}}1'$ and ${\rm{C}}2'$. To address this problem, we further introduce the auxiliary variables ${\lambda _k} = {\beta _k}{\theta _k}$, substitute them into problem ${{\cal P}_{1.1}}$ and have
\begin{subequations}
\begin{align}
&{{\cal P}_{1.2}}:\;\;\;\;\mathop {\min }\limits_{{\tau _0},{\bm{\tau}},{\bm{\lambda }},{\theta_0},{\bm{\theta}}} \;\sum\limits_{i = 0}^K {\theta_i} \\
{\rm{s}}{\rm{.t}}{\rm{.}}\;\;&{\rm{C}}1'': W{\tau _k}{\log _2}\left( {1 + \frac{{\xi {\lambda _k}{h_k}{g_k}}}{{{\tau _k}W{N_0}}}} \right) \ge R_k^{\min },\;\forall k,\\ \notag
&{\rm{C}}2'':\left( {\frac{{\frac{{a\left( {{\theta _k} - {\lambda _k}} \right)}}{{{\tau _k}}}{h_k} + d}}{{\frac{{{\theta _k} - {\lambda _k}}}{{{\tau _k}}}{h_k} + v}} - \frac{d}{v}} \right){\tau _k}\\
&\;+\sum\limits_{i = 0,i \ne k}^K {\left( {\frac{{a\frac{{{\theta _i}}}{{{\tau _i}}}{h_k} + d}}{{{{\frac{{{\theta _i}}}{{{\tau _i}}}}}{h_k} + v}} - \frac{d}{v}} \right){\tau _i}}  \ge {p_{c,k}}{\tau _k},\forall k,\\
&{\rm{C}}4':0 \le {\lambda _k} \le {\theta _k},\forall k\\
&{\rm{C}}3,\;{\rm{C}}5'.\notag
\end{align}
\end{subequations}

{\bf{Theorem 1.}} Problem ${{\cal P}_{1.2}}$  is convex.

\emph{Proof.} Please refer to Appendix A.

Although problem ${{\cal P}_{1.2}}$ is convex, existing numerical convex program solvers, e.g, CVX tool, cannot directly handle constraint ${\rm{C}}2''$ in (9c) because the CVX tool treats ${\rm{C}}2''$ as a non-convex constraint. Although the successive convex approximation method can be applied on ${\rm{C}}2''$ so that CVX can be employed, such an approach may not be able to provide useful insights into the energy minimization based resource allocation scheme. In view of the above, we leverage the KKT conditions to derive closed-form expressions for some of the optimization variables, thus providing a better understanding of the investigated problem, and devise an iterative algorithm to solve problem ${{\cal P}_{1.2}}$.

\subsection{Problem Solution}
\newcounter{mytempeqncnt}
\begin{figure*}[t]
\normalsize
\setcounter{mytempeqncnt}{\value{equation}}
\begin{align}\notag \label{10}
\mathcal{L} &= \sum\limits_{i = 0}^K {{\theta _i}} +\sum\limits_{k = 1}^K {{\alpha _k}\left[ {{p_{c,k}}{\tau _k} - \left( {\frac{{\frac{{a\left( {{\theta _k} - {\lambda _k}} \right)}}{{{\tau _k}}}{h_k} + d}}{{\frac{{{\theta _k} - {\lambda _k}}}{{{\tau _k}}}{h_k} + v}} - \frac{d}{v}} \right){\tau _k} - \sum\limits_{i = 0,i \ne k}^K {\left( {\frac{{a\frac{{{\theta _i}}}{{{\tau _i}}}{h_k} + d}}{{{{\frac{{{\theta _i}}}{{{\tau _i}}}}}{h_k} + v}} - \frac{d}{v}} \right){\tau _i}} } \right]}\\
&+\sum\limits_{k = 1}^K {{\varepsilon _k}\left[ {R_k^{\min } - W{\tau _k}{{\log }_2}\left( {1 + \frac{{\xi {\lambda _k}{h_k}{g_k}}}{{{\tau _k}W{N_0}}}} \right)} \right]}+\vartheta \left[ {\sum\limits_{i = 0}^K {{\tau _i}}  - T} \right]{\rm{ + }}\sum\limits_{k = 1}^K {{\kappa _k}\left( {{\lambda _k} - {\theta _k}} \right)} {\rm{ + }}\sum\limits_{i = 0}^K {{\omega _i}\left( {{\theta _i} - P_{\rm{max}}{\tau _i}} \right)}
\end{align}
\hrulefill
\end{figure*}
The partial Lagrangian function for problem ${{\cal P}_{1.2}}$ is written in \eqref{10}, as shown at the top of the next page, where $\bm{\alpha}=\left[ {{\alpha _1},{\alpha _2}, \ldots ,{\alpha _K}} \right]\succeq\bm{0}$, $\bm{\varepsilon}=\left[ {{\varepsilon _1},{\varepsilon _2}, \ldots ,{\varepsilon _K}} \right]\succeq\bm{0}$, $\bm{\kappa}=\left[ {{\kappa _1},{\kappa _2}, \ldots ,{\kappa _K}} \right]\succeq\bm{0}$, and $\bm{\omega}=\left[ {{\omega _0},{\omega _1}, \ldots ,{\omega _K}} \right]\succeq\bm{0}$ are Lagrange multiplier vectors corresponding to constraints ${\rm{C}}2''$, ${\rm{C}}1''$,  ${\rm{C}}4$, and ${\rm{C}}5'$, respectively, and $\vartheta\geq 0$ is the Lagrange multiplier associated with constraint ${\rm{C}}3$.

Taking the partial derivative of ${\cal L}$ with respect to each optimization variable, respectively, we can obtain the following results,
\begin{align}\label{11}
\frac{{\partial {\cal L}}}{{\partial {\theta _0}}} = 1 - \sum\limits_{k = 1}^K {\frac{{{\alpha _k}\left( {av - d} \right){h_k}}}{{{{\left( {\frac{{{\theta _0}}}{{{\tau _0}}}{{h_k}} + v} \right)}^2}}}}  + {\omega _0},
\end{align}
\begin{align}\label{12}
\frac{{\partial {\cal L}}}{{\partial {\theta _k}}} = 1 - \frac{{{\alpha _k}\left( {av - d} \right){h_k}}}{{{{\left( {\frac{{{\theta _k} - {\lambda _k}}}{{{\tau _k}}}{h_k} + v} \right)}^2}}} - {\kappa _k} + {\omega _k}, \forall k,
\end{align}
\begin{align}\label{13}
\frac{{\partial {\cal L}}}{{\partial {\lambda _k}}}{\rm{ = }}{\kappa _k} + \frac{{{\alpha _k}\left( {av - d} \right){h_k}}}{{{{\left( {\frac{{{\theta _k} - {\lambda _k}}}{{{\tau _k}}}{h_k} + v} \right)}^2}}} - \frac{{{\varepsilon _k}W\xi {h_k}{g_k}}}{{\left( {W{N_0} + \xi {h_k}{g_k}\frac{{{\lambda _k}}}{{{\tau _k}}}} \right)\ln 2}},
\end{align}
\begin{align}\label{14}\notag
\frac{{\partial {\cal L}}}{{\partial {\tau _0}}} =  & - \sum\limits_{k = 1}^K {{\alpha _k}\left( {a - \frac{d}{v} + \frac{{\left( {d - av} \right)\left( {\frac{{2{\theta _0}}}{{{\tau _0}}}{h_k} + v} \right)}}{{{{\left( {\frac{{{\theta _0}}}{{{\tau _0}}}{h_k} + v} \right)}^2}}}} \right)}  \\
&  + \vartheta- {\omega _0}{P_{\max }},
\end{align}
\begin{align}\label{15}\notag
\frac{{\partial {\cal L}}}{{\partial {\tau _k}}} &= {\alpha _k}{p_{c,k}} - {\alpha _k}\left( {a - \frac{d}{v} + \frac{{\left( {d - av} \right)\left( {\frac{{2\left( {{\theta _k} - {\lambda _k}} \right)}}{{{\tau _k}}}{h_k} + v} \right)}}{{{{\left( {\frac{{{\theta _k} - {\lambda _k}}}{{{\tau _k}}}{h_k} + v} \right)}^2}}}} \right)  \\ \notag
 &+ \vartheta- {\omega _k}{P_{\max }} - {\varepsilon _k}W{\log _2}\left( {1 + \frac{{\xi {\lambda _k}{h_k}{g_k}}}{{{\tau _k}W{N_0}}}} \right) \\
 & + \frac{{{\varepsilon _k}W\xi {h_k}{g_k}\frac{{{\lambda _k}}}{{{\tau _k}}}}}{{\left( {W{N_0} + \xi {h_k}{g_k}\frac{{{\lambda _k}}}{{{\tau _k}}}} \right)\ln 2}}.
\end{align}

By solving $\frac{{\partial {\cal L}}}{{\partial {\theta _0}}}=0$ for $\theta_0 / \tau_0$, the optimal PB transmission power in time slot $\tau_0$ is given by
\begin{align}\label{16}
P_0^* = \frac{{{\theta _0}}}{{{\tau _0}}} = {\varphi ^{ - 1}}\left( {1 + {\omega _0}} \right),
\end{align}
where $\varphi \left( x \right) = \sum\limits_{k = 1}^K {\frac{{{\alpha _k}\left( {av - d} \right){h_k}}}{{{{\left( {{h_k}x + v} \right)}^2}}}} $ and ${\varphi ^{ - 1}}\left( x \right)$ denotes its inverse function.  As we have shown ${av - d}>0$ in Appendix A,  $\varphi \left( x \right)$ decreases with $x$ for $x\ge 0$ and we can find a unique  $\frac{{{\theta _0}}}{{{\tau _0}}}\ge 0$ satisfying \eqref{16} via one-dimensional searching methods.

%
%

By letting $\frac{{\partial {\cal L}}}{{\partial {\theta _k}}}=0$ and solving it for $(\theta_k - \lambda_k)/ \tau_k$, we have
\begin{align}\notag \label{17}
P_k^*\left( {1 - \beta _k^*} \right) &= \frac{{{\theta _k} - {\lambda _k}}}{{{\tau _k}}} \\
&= {\left[ {\frac{1}{{{h_k}}}\sqrt {\frac{{{\alpha _k}\left( {av - d} \right){h_k}}}{{1 + {\omega _k} - {\kappa _k}}}}  - \frac{v}{{{h_k}}}} \right]^ + },
\end{align}
where ${\left[ x \right]^ + } = \max \left\{ {0,x} \right\}$.

Substituting  \eqref{17} into \eqref{13} and solving $\frac{{\partial {\cal L}}}{{\partial {\lambda _k}}}=0$ for $\lambda_k/ \tau_k$, we obtain
\begin{align} \label{18}
P_k^*\beta_k^*=\frac{{{\lambda _k}}}{{{\tau _k}}} = {\left[ {\frac{{{\varepsilon _k}W}}{{\left( {{\kappa _k} + \left( {1 + {\omega _k} - {\kappa _k}} \right)\ln 2} \right)}} - \frac{{W{N_0}}}{{\xi {h_k}{g_k}}}} \right]^ + }.
\end{align}
One can see from \eqref{18} that  $P_k^*\beta_k^*$ increases with ${h_k}{g_k}$. This indicates that in order to minimize the energy consumed by the PB, the IoT node with high channel gains from the PB and to the IF should backscatter a large portion of the incident signal power to the IF.

Based on \eqref{17} and \eqref{18}, the optimal $P_k$ is calculated as
\begin{align}\label{19a}\notag
P_k^* &= {\left[ {\frac{1}{{{h_k}}}\sqrt {\frac{{{\alpha _k}\left( {av - d} \right){h_k}}}{{1 + {\omega _k} - {\kappa _k}}}}  - \frac{v}{{{h_k}}}} \right]^ + }\\
&+{\left[ {\frac{{{\varepsilon _k}W}}{{\left( {{\kappa _k} + \left( {1 + {\omega _k} - {\kappa _k}} \right)\ln 2} \right)}} - \frac{{W{N_0}}}{{\xi {h_k}{g_k}}}} \right]^ + },
\end{align}
and the optimal $\beta_k$ is given by
\begin{align}
\beta_k^*= \frac{1}
{\frac{{\left[ {\frac{1}{{{h_k}}}\sqrt {\frac{{{\alpha _k}\left( {av - d} \right){h_k}}}{{1 + {\omega _k} - {\kappa _k}}}}  - \frac{v}{{{h_k}}}} \right]^ + }}{{\left[ {\frac{{{\varepsilon _k}W}}{{\left( {{\kappa _k} + \left( {1 + {\omega _k} - {\kappa _k}} \right)\ln 2} \right)}} - \frac{{W{N_0}}}{{\xi {h_k}{g_k}}}} \right]^ + }}+1}.
\end{align}

Substituting \eqref{16} into \eqref{14}, we can rewrite $\frac{{\partial {\cal L}}}{{\partial {\tau _0}}}$ as
\begin{align}\notag \label{19}
&\frac{{{\partial {\cal L}}}}{{\partial {\tau _0}}} = \vartheta  - {\omega _0}{P_{\max }} - \sum\limits_{k = 1}^K {\alpha _k}\\
&\times \left( {a - \frac{d}{v} + \frac{{\left( {d - av} \right)\left( {2{\varphi ^{ - 1}}\left( {1 + {\omega _0}} \right){h_k} + v} \right)}}{{{{\left( {{\varphi ^{ - 1}}\left( {1 + {\omega _0}} \right){h_k} + v} \right)}^2}}}} \right).
\end{align}
One can observe from (21) that the Lagrangian function (10) is a linear function of ${\tau _0}$. By substituting \eqref{17} and \eqref{18} into \eqref{15}, we find that $\frac{{\partial {\cal L}}}{{\partial {\tau _k}}} $ is free of $\tau_k$, thus the Lagrangian function (10) is also a linear function of ${\tau _k}$. The above observations indicate that the optimal $\tau_0$ and $\tau_k$ can be obtained by solving the following linear programming problem, which is obtained by substituting \eqref{16}-\eqref{19a} into ${{\cal P}_{1.2}}$.
\begin{subequations}
\begin{align}
&{{\cal P}_{1.3}}:\;\;\;\;\mathop {\min }\limits_{{\tau _0},{\bm{\tau}}} \;\sum\limits_{i = 0}^K {P^*_i\tau_i} \\
{\rm{s}}{\rm{.t}}{\rm{.}}\;\;&{\rm{C}}1''': W{\tau _k}{\log _2}\left( {1 + \frac{{\xi {\beta^* _k}{h_k}{g_k}}}{{W{N_0}}}} \right) \ge R_k^{\min },\;\forall k,\\ \notag
&{\rm{C}}2''':\left( {\frac{{aP_k^*\left( {1 - \beta _k^*} \right){h_k} + d}}{{P_k^*\left( {1 - \beta _k^*} \right){h_k} + v}} - \frac{d}{v}} \right){\tau _k}\\
&\;+\sum\limits_{i = 0,i \ne k}^K {\left( {\frac{{aP_i^*{h_k} + d}}{{P_i^*{h_k} + v}} - \frac{d}{v}} \right){\tau _i}}  \ge {p_{c,k}}{\tau _k},\forall k,\\
&{\rm{C}}3.\notag
\end{align}
\end{subequations}

Having obtained $\{{{\tau _0},{\bm{\tau}},{\bm{\lambda }},{\theta_0},{\bm{\theta}}}\}$ for given Lagrange multipliers, the Lagrange multipliers can be updated in an iterative manner by using the gradient method as follows,
\begin{align}\notag
\alpha _k^{(n + 1)} &= \alpha _k^{(n)} - {\ell _1}\left( {p_{c,k}}{\tau _k} - \bigg( {\frac{{\frac{{a\left( {{\theta _k} - {\lambda _k}} \right)}}{{{\tau _k}}}{h_k} + d}}{{\frac{{{\theta _k} - {\lambda _k}}}{{{\tau _k}}}{h_k} + v}} - \frac{d}{v}} \right){\tau _k}\\
 &- \sum\limits_{i = 0,i \ne k}^K {\left( {\frac{{a\frac{{{\theta _i}}}{{{\tau _i}}}{h_k} + d}}{{\frac{{{\theta _i}}}{{{\tau _i}}}{h_k} + v}} - \frac{d}{v}} \right){\tau _i}}  \bigg), \forall k,
\end{align}
\begin{align}
\varepsilon_k^{(n + 1)} \!=\! \varepsilon _k^{(n)} \!-\! {\ell _2}\left(\!\! {R_k^{\min } \!-\! W{\tau _k}{{\log }_2}\left( {1\! +\! \frac{{\xi {\lambda _k}{h_k}{g_k}}}{{{\tau _k}W{N_0}}}}\!\! \right)} \right), \forall k,
\end{align}
\begin{align}
\omega _i^{(n + 1)} = \omega _i^{(n)} - {\ell _3}\left( {{\theta _i} - {P_{{\rm{max}}}}{\tau _i}} \right), i = \left\{ {0,1,...,K} \right\},
\end{align}
\begin{align}
\vartheta _{}^{(n + 1)} = \vartheta _{}^{(n)} - {\ell _4}\left( {\sum\limits_{i = 0}^K {{\tau _i}}  - T} \right),
\end{align}
\begin{align}
\kappa _k^{(n + 1)} = \kappa _k^{(n)} - {\ell _5}\left( {{\lambda _k} - {\theta _k}} \right), \forall k,
\end{align}
where $n \ge 0$ is the iteration index, and ${\ell _1}$, ${\ell _2}$, ${\ell _3}$, ${\ell _4}$, and ${\ell _5}$ are the  step sizes. To guarantee the convergence of the gradient method, the values of the step sizes can be set following \cite{boyd2004convex}.

The above Lagrange dual method for solving P1 is summarized in Algorithm 1, where $\phi$ is the number of iteration.
\renewcommand{\algorithmicrequire}{\textbf{Input:}}
\renewcommand{\algorithmicensure}{\textbf{Output:}}
\begin{algorithm}
\caption{The dynamic PB transmit power scheme}
\label{alg1}
\begin{algorithmic}[1]
\REQUIRE $K$ IoT nodes.
\ENSURE $\tau_0^*, \tau_k^*, \beta_k^*, P_0^*, P_k^*$.
\renewcommand{\algorithmicensure}{\textbf{Initialize:}}
\ENSURE $\phi=0$
\STATE Initialize $\bm{\alpha}(\phi), \bm{\varepsilon}(\phi),\bm{\kappa}(\phi), \bm{\omega}(\phi), \vartheta(\phi)$.
\REPEAT
\STATE Obtain $P_k^*$ and $\beta_k^*$ from (19) and (20), respectively.
\STATE Obtain $\tau_0^*$ and $\tau_k^*$ by solving ${\cal P}_{1.3}$ with CVX.
\STATE $\phi=\phi+1$;
\STATE Update $\bm{\alpha}(\phi), \bm{\varepsilon}(\phi),\bm{\kappa}(\phi), \bm{\omega}(\phi), \vartheta(\phi)$ using (23)-(27).
\UNTIL The values of Lagrange multipliers converge.
\IF {$0 < P^*_0, P^*_k \leq P_{max}$}
\STATE return $\tau^*_0$, $\tau^*_k$, $\beta^*_k, P^*_0, P^*_k$;
\ELSE
\STATE return $\tau^*_0 = 0, \tau^*_k = 0, \beta^*_k = 0, P^*_0 = 0, P^*_k = 0$;
\ENDIF
\end{algorithmic}
\end{algorithm}

\subsection{Insights}
By applying the KKT approach to ${{\cal P}_{1.2}}$, we obtain the following two key insights into PB energy minimization based resource allocation scheme

\textbf{Insight 1.} The energy consumed by the PB in a time block is minimized when at least one IoT node consumes all the harvested energy while maintaining its minimum throughput at the minimum required level.

\emph{Proof. } Please refer to Appendix B.

\textbf{Insight 2.} The PB's energy consumption is minimized when the whole time block is used up for EH and BackCom by all the IoT nodes.

\emph{Proof. } Please refer to Appendix C.

\textbf{Remark 1.} We can explain \textbf{Insight 2} as follows. Since the dynamic PB transmit power scheme allows the PB to use the lowest possible transmit power in each time slot according to the channel conditions from the PB to the corresponding IoT node and from the IoT node to the IF, it allows each IoT node to use the longest possible  time duration to perform EH and BackCom to meet their throughput requirement. As a result, the whole time block is used up.

\section{PB Energy Consumption Minimization with Static Transmit Power}
In this section, to reduce the implementation complexity  required for adjusting the PB’s transmit power every time slot, we introduce a static PB transmit power scheme, where the PB transmit power is kept constant in each time block. Accordingly, we formulate another joint optimization problem to minimize the PB energy consumption and solve the it by employing the BCD method and the SCA technique.

\subsection{Problem Formulation}
Under the condition that the PB transmit power is constant in each time block, we minimize the energy consumption of the PB by optimizing the PB transmit power, the time slot duration and the backscatter reflection coefficient of each IoT node, subject to the minimum throughput requirement of each IoT node. The optimization problem is formulated as
\begin{subequations}
\begin{align}
&{{\cal P}_2}:\;\;\;\;\mathop {\min }\limits_{{\tau _0},{\bm{\tau}},{\bm{\beta }},{P}} \;\sum\limits_{i = 0}^K {{P}{\tau _i}} \\
{\rm{s}}{\rm{.t}}{\rm{.}}\;\;&{\rm{C}}1-1:\; W{\tau _k}{\log _2}\left( {1 + \frac{{\xi\beta_k {P}{h_k}{g_k}}}{{W{N_0}}}} \right) \ge R_k^{\min },\;\forall k,\\ \notag
&{\rm{C}}2-1:\left( {\frac{{a\left( {1 - {\beta _k}} \right){P}{h_k} + d}}{{\left( {1 - {\beta _k}} \right){P}{h_k} + v}} - \frac{d}{v}} \right){\tau _k}\\
&\;\;\;\;\;+\sum\limits_{i = 0,i \ne k}^K {\left( {\frac{{a{P}{h_k} + d}}{{{P}{h_k} + v}} - \frac{d}{v}} \right){\tau _i}} \ge {p_{c,k}}{\tau _k},\forall k,\\
&{\rm{C}}3; {\rm{C}}4,\notag\\
&{\rm{C}}5-1:0 < P \le {P_{\max }},
\end{align}
\end{subequations}
where $P$ denotes the PB transmit power in the current time block.

By observing ${{\cal P}_2}$, we can see that replacing $P\tau_i$ by an auxiliary variable, e.g., $P\tau_i = \theta_i$, cannot transform ${{\cal P}_2}$ into a convex form and the reason is as follows. Since the PB transmit power $P$ is fixed across different time slots in a time block, the above auxiliary variable will lead to another constraint, i.e., $\theta_k/\tau_k=P, \forall k$, which is still non-convex and hard to solve. Thus, in the following, we employ the BCD method and SCA technique to solve ${{\cal P}_2}$.

\subsection{Problem Transformation and Solution}
For a given $P$, we can solve ${{\cal P}_2}$ by employing the BCD method to obtain the sub-optimal solution of the time slot duration and the backscatter reflection coefficient of each IoT node. The obtained solutions can then be substituted into $P_2$ to find the sub-optimal solution of $P$. Thus, ${{\cal P}_2}$ can be solved in an iterative manner. Next, we introduce the BCD method in details.

For a given $P$, we introduce the auxiliary variables $L_k=\beta_k\tau_k, \forall k$, substitute them into ${{\cal P}_2}$ and obtain
\begin{subequations}
\begin{align}
&{{\cal P}_{2.1}}:\;\;\;\;\mathop {\min }\limits_{{\tau _0},{\bm{\tau}},{\bm{L }}} \;\sum\limits_{i = 0}^K {{P}{\tau _i}} \\
{\rm{s}}{\rm{.t}}{\rm{.}}\;\;&{\rm{C}}1-2:\; W{\tau _k}{\log _2}\left( {1 + \frac{{\xi L_k {P}{h_k}{g_k}}}{{W\tau_k{N_0}}}} \right) \ge R_k^{\min },\;\forall k,\\
&{\rm{C}}2-2: E_k^{\rm{total}'}(\frac{L_k}{\tau_k}) \ge {p_{c,k}}{\tau _k},\forall k,\\
&{\rm{C}}3,\notag\\
&{\rm{C}}4-1:0 < L_k \le \tau_k, \forall k,
\end{align}
\end{subequations}
where ${\bm{L}} = \left[ {{L_1},{L_2}, \cdots ,{L_K}} \right]$, $E_k^{\rm{total}'}(\frac{L_k}{\tau_k})=\left( {\frac{{a\left( {1 - {\frac{L_k}{\tau_k}}} \right){P}{h_k} + d}}{{\left( {1 - {\frac{L_k}{\tau_k}}} \right){P}{h_k} + v}} - \frac{d}{v}} \right){\tau _k}+\sum\limits_{i = 0,i \ne k}^K {\left( {\frac{{a{P}{h_k} + d}}{{{P}{h_k} + v}} - \frac{d}{v}} \right){\tau _i}}$.

Following similar steps as in Appendix A, we can prove that ${\cal P}_{2.1}$ is a convex problem, but we cannot directly use CVX due to constraint $\rm{C}2-2$ even though it is convex. Thus, we apply the SCA technique to convert $\rm{C}2-2$ to a linear form based on $\textbf{Lemma\;1}$. \\

$\textbf{Lemma\;1:}$ Letting $Y_k^{(j)}=(\frac{L_k}{\tau_k})^j, j\ge1$, denote the obtained solution of $Y_k$ after the $j$th iteration, then it satisfies that \cite{8901136}
\begin{equation}
E_k^{\rm{total}'}(\frac{L_k}{\tau_k})\ge E_k^{\rm{total}'}(Y_k^{(j)}), \forall k,
\end{equation}
where
\begin{equation}
\begin{split}
E_k^{\rm{total}'}(Y_k^{(j)})&=\left( {\frac{{a\left( {1 - {Y_k^{(j)}}} \right){P}{h_k} + d}}{{\left( {1 - {Y_k^{(j)}}} \right){P}{h_k} + v}} - \frac{d}{v}} \right){\tau _k}\\
&+\frac{Ph_kd-aPh_kv}{(Ph_k-Y_k^{(j)}Ph_k+v)^2}(L_k-Y_k^{(j)}\tau_k)\\
&+\sum\limits_{i = 0,i \ne k}^K {\left( {\frac{{a{P}{h_k} + d}}{{{P}{h_k} + v}} - \frac{d}{v}} \right){\tau _i}},
\end{split}
\end{equation}
and the equalities in (30) only hold when $\frac{L_k}{\tau_k}=Y_k^{(j)}$.

\emph{Proof. } Please refer to Appendix D.

By substituting (31) into ${\cal P}_{2.1}$, ${\cal P}_{2.1}$ is equivalently transformed into \begin{subequations}
\begin{align}
&{{\cal P}_{2.1}^{'}}: \{\tau_0^*,\bm{\tau^*},\bm{L^*}\}= {\rm{arg}} \mathop {\min }\limits_{{\tau _0},{\bm{\tau}},{\bm{L }}} \;\sum\limits_{i = 0}^K {{P}{\tau _i}} \\
{\rm{s}}{\rm{.t}}{\rm{.}}\;\;&{\rm{C}}1-2;\notag\\
&{\rm{C}}2-3: E_k^{\rm{total}'}(Y_k^{(j)}) \ge {p_{c,k}}{\tau _k},\forall k,\\
&{\rm{C}}3;{\rm{C}}4-1.\notag
\end{align}
\end{subequations}

Note that ${\cal P}_{2.1}^{'}$ can be directly solved by CVX.

Substituting the obtained $\tau_0^*, \bm{\tau^*}$ and $\bm{L^*}$ into ${\cal P}_{2}$, we have
\begin{subequations}
\begin{align}
&{{\cal P}_{2.2}}:\;\;\;\;\mathop {\min }\limits_{{P}} \;\sum\limits_{i = 0}^K {{P}{\tau _i}} \\
{\rm{s}}{\rm{.t}}{\rm{.}}\;\;&{\rm{C}}1-1; {\rm{C}}2-1; {\rm{C}}5-1.\notag\;
\end{align}
\end{subequations}

According to Appendix A, ${\cal P}_{2.2}$ is also a convex problem, and we use the same SCA method for solving ${\cal P}_{2.1}$ to transform $\rm{C}2-1$ into a linear form, then ${\cal P}_{2.2}$ can be directly solved by CVX. The transformation and solution process is omitted for brevity.

The above BCD method and SCA technique-based procedure for solving ${{\cal P}_2}$ is summarized in Algorithm 2, where $\epsilon$ denotes the number of iteration.

\begin{algorithm}
\caption{The static PB transmit power scheme}
\label{alg1}
\renewcommand{\algorithmicrequire}{\textbf{Input:}}
\renewcommand{\algorithmicensure}{\textbf{Output:}}
\begin{algorithmic}[1]
\STATE  $K$ IoT nodes.
\STATE  $\tau_0^*, \tau_k^*, \beta_k^*, P^*$.
\renewcommand{\algorithmicensure}{\textbf{Initialize:}}
\STATE  $\epsilon=0$
\STATE Initialize $P(\epsilon)$.
\REPEAT
\STATE Obtain $\tau_0^{*}(\epsilon), \tau_k^{*}(\epsilon), L_k^{*}(\epsilon)$ by solving ${\cal P}_{2.1}$ with SCA method and CVX.
\STATE Obtain $\beta_k^{*}(\epsilon)=\frac{L_k^{*}(\epsilon)}{\tau_k^{*}(\epsilon)}$.
\STATE $\epsilon=\epsilon+1$.
\STATE Obtain $P(\epsilon)$ by solving ${\cal P}_{2.2}$ with SCA method and CVX.
\UNTIL $\tau_0^{*}(\epsilon), \tau_k^{*}(\epsilon), L_k^{*}(\epsilon)$ converge.
\IF{$0 < P(\epsilon) \leq P_{max}$}
\STATE return $\tau^*_0, \tau^*_k, \beta^*_k, P^* = P(\epsilon);$%
\ELSE
\STATE return $\tau^*_0 = 0, \tau^*_k = 0, \beta^*_k = 0, P^* = 0;$
\ENDIF
\end{algorithmic}
\end{algorithm}

\subsection{Insights}
\textbf{Insight 3.}
Different from the dynamic PB transmit power scheme, for the static PB transmit power scheme, we find that minimizing the energy consumption of the PB does not require the whole time block to be used up for EH and BackCom by all the IoT nodes.

\emph{Proof. } Please refer to Appendix E.

\textbf{Remark 2.}We can explain Insight 3 as follows. Since the PB transmit power keeps constant during the whole time block, the PB energy consumption is minimized by allocating each IoT node with the minimum required time durations for EH and BackCom that meet their throughput requirement. Thus, the whole time block may not be used up.

\section{Convergence and Run Time Analysis}
We first analyze the convergence of Algorithms 1 \& 2. In Algorithm 1, i.e., the dynamic PB transmit power scheme, where the Lagrange dual method is used to solve ${\cal P}_{1.3}$, since it is a convex problem, the Lagrange multipliers $\bm{\alpha}, \bm{\varepsilon},\bm{\kappa}, \bm{\omega}, \vartheta$ are guaranteed to converge to the optimal solution of ${\cal P}_{1.3}$. In Algorithm 2, i.e., the static PB transmit power scheme, after each iteration of the BCD process, the objective function value of
${\cal {P}}_{2}^{'}$ is non-increasing with updated variables. Meanwhile, ${\cal P}_{2}$ is lower bounded by its constraints. Thus, the BCD method is guaranteed to converge to the locally optimal solutions of ${\cal P}_{2}$. Since a convex function is globally lower bounded by its first-order Taylor expansion as given in (30), the achieved value of the objective function is non-increasing after each iteration in ${\cal {P}}_{2.1}^{'}$ and ${\cal {P}}_{2.1}^{'}$ is lower bounded by its constraints. Thus, the SCA technique is guaranteed to converge to the locally optimal solutions of ${\cal {P}}_{2.1}^{'}$ \cite{8901136}.

Next, we evaluate the run time of the proposed Algorithms 1 \& 2 through simulations, which are performed using MATLAB 2018 and a laptop with the following configurations: Intel(R) Core(TM) i7-9750H CPU @ 2.6GHZ, RAM 16 GB. In Algorithm 1, the number of iterations for updating the Lagrange multipliers, which is denoted by $\Delta_1$, is around 3-4, and each iteration spends around 1.33s. Accordingly, the run time of Algorithm 1 is around 1.33$\Delta_1$s. In Algorithm 2, we denote the number of iterations of the BCD process by $\Delta_2$, and the number of iterations of the SCA process for solving  ${\cal P}_{2.1}$ and ${\cal P}_{2.2}$ as $\Delta_3$ and $\Delta_4$, respectively. According to the simulations, the run time of Algorithm 2 is around $\Delta_2(1.284\Delta_3+1.296\Delta_4)$s, where $\Delta_2$, $\Delta_3$ and $\Delta_4$ are in the value ranges of 3-4, 2-3 and 2-3, respectively.

\section{Numerical Results Analysis}
In this section, we present numerical results to evaluate the performance of the proposed dynamic PB transmit power scheme and static PB transmit power scheme as compared with two benchmark schemes, which are the EE maximization scheme \cite{9562293} and the throughput maximization scheme\cite{8901136}. We assume path loss with distance exponent of $\alpha$ as large scale fading and consider Rayleigh fading following $exp(1)$ as small scale fading. The values of the simulation parameters are listed in {\bf{Table 1}} unless otherwise specified. .

\begin{table}[h]
\centering
\begin{tabular}{|l|l|}
\multicolumn{2}{c}{\textbf{Table 1 Simulation Parameters}}\\
\hline
Parameter  & Value  \\
\hline
Number of IoT nodes $\emph{K}$ & 5 \\
\hline
Time block $T$ & 10s\\
\hline
Non-linear EH model parameters $a, d, v$ & 2.463, 1.635, 0.826 \cite{9405302}\\
\hline
Channel bandwidth $W$ & 400kHz \cite{7999248}\\
\hline
Noise power spectral density $N_0$ & $-$110 dBm/Hz \cite{9405302}\\
\hline
Path loss exponent $\alpha$ & 3\\
\hline
Maximum PB transmit power $P_{max}$ & 23 dBm\\
\hline
Circuit power consumption of &  \multirow{2}{*}{200uw}\\ each IoT node $P_{c,k}$ & \\
\hline
PB-IF distance $r$ & 25m\\
\hline
Minimum throughput requirement of &  \multirow{2}{*}{2400bit/s}\\ each IoT node $R^{min}$ & \\
\hline
\end{tabular}
\end{table}


\begin{figure}[h]
\centerline{\includegraphics[width=3.5in]{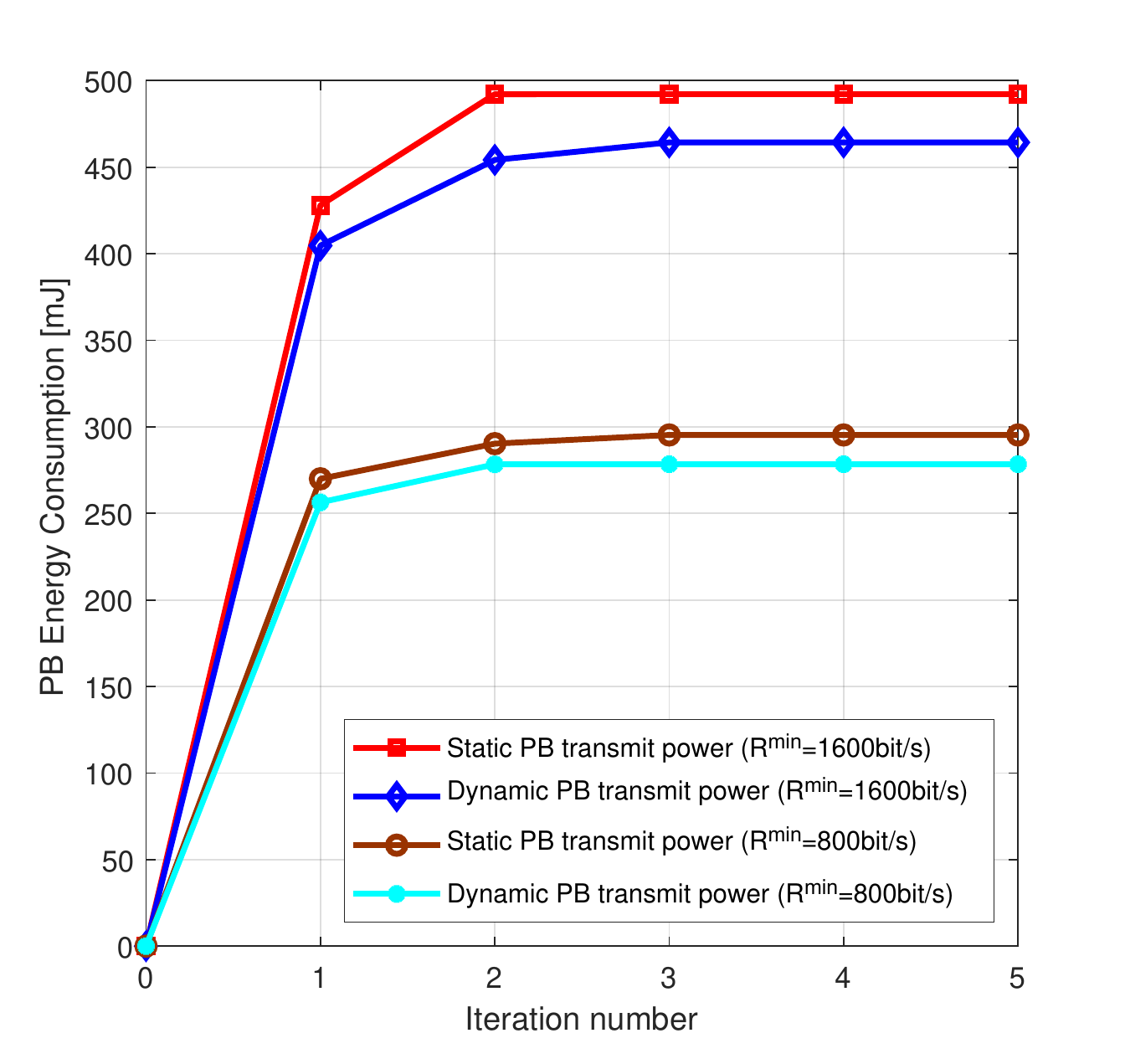}}
\caption{PB energy consumption versus iteration number $(r=25m)$}
\end{figure}
Fig. 2 shows the PB's energy consumption achieved by the proposed Algorithm 1 (the dynamic PB transmit power scheme) and Algorithm 2 (the static PB transmit power scheme) versus the iteration. We can see that both algorithms converge very fast and that the PB energy consumption of the dynamic PB transmit power scheme is lower than that of the static PB transmit power scheme. This is because in the dynamic scheme, the minimum possible PB transmit power is used in each time slot according to the channel condition from the PB to the corresponding IoT node and from the IoT node to the IF; while in the static scheme, the PB transmit power needs to be high enough to ensure that the IoT node throughput requirement is met even for the IoT node that sees the worst channel condition. Furthermore, the gap between the PB energy consumption achieved by the two schemes increases with the minimum throughput requirement $R^{\rm{min}}$. This indicates that
the performance advantage of the dynamic PB transmit power scheme over the static PB transmit power scheme increases with $R^{\rm{min}}$.


\begin{figure}[h]
\centerline{\includegraphics[width=3.5in]{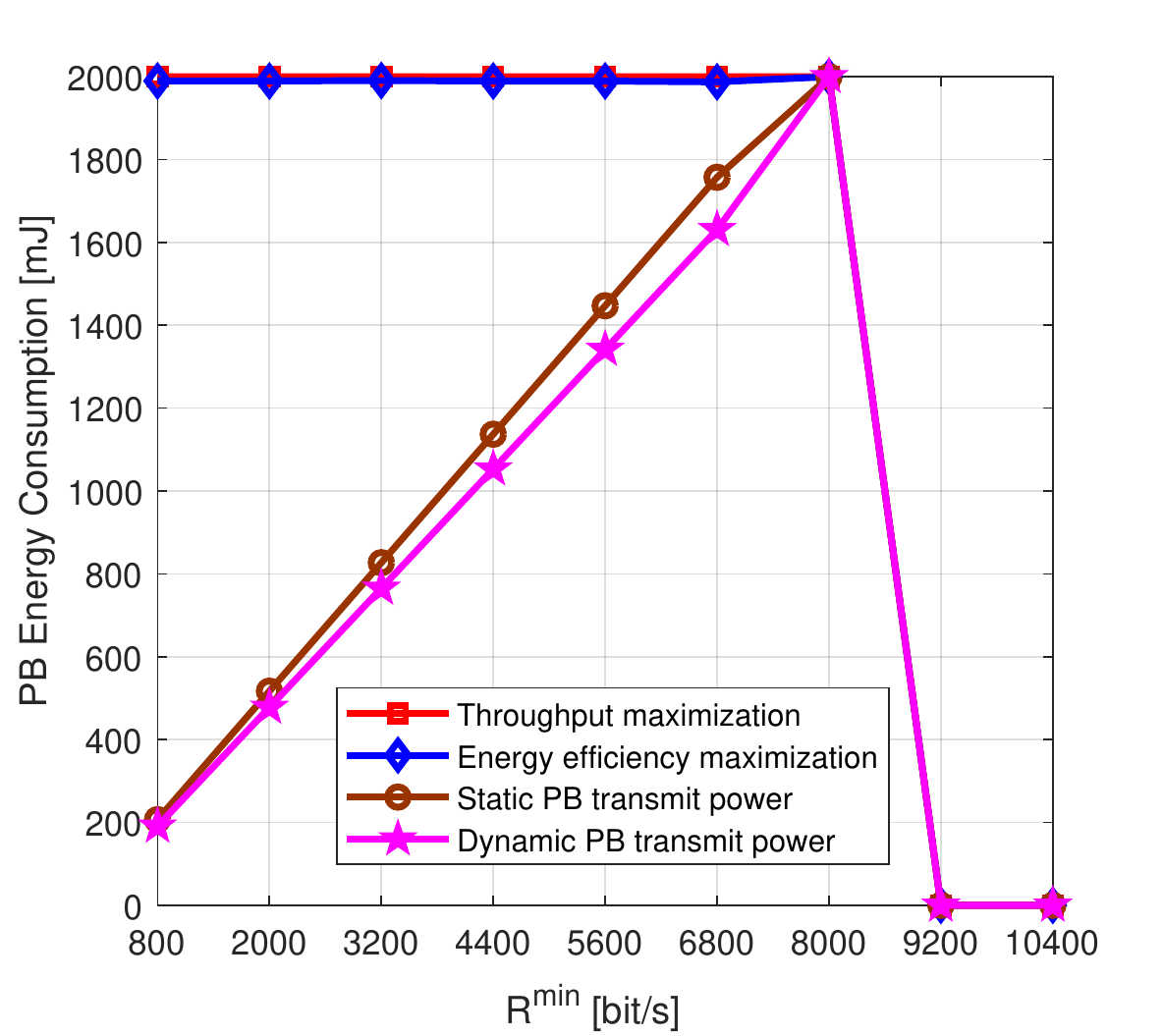}}
\caption{PB energy consumption versus the minimum required throughput $R^{\rm{min}}$ $(r=25m)$}
\end{figure}
Fig. 3 plots the PB energy consumption versus the minimum throughput requirement $R^{\rm{min}}$ of the IoT nodes under four different schemes. For both the throughput maximization scheme \cite{8901136} and the EE maximization scheme \cite{9562293}, the maximum possible amount of energy is consumed by the PB in a time block for $R^{min}$ values up to 8000 bit/s, because both throughput maximization and EE maximization lead to the use of the maximum allowed PB transmit power. For the proposed dynamic PB transmit power scheme and static PB transmit power scheme, the PB energy consumption increases with $R^{\rm{min}}$ before it reaches 8000 bit/s, because they both use the minimum possible PB transmit power and transmission time, which increases with the higher throughput requirement. For all the four considered schemes, when $R^{\rm{min}}$ exceeds 8000 bit/s, the PB energy consumption drops to 0 since none of them can find a feasible PB transmit power to satisfy the high throughput requirement and the schemes return a zero transmit power of the PB (see Algorithms 1 \& 2).

\begin{figure}[h]
\centerline{\includegraphics[width=3.5in]{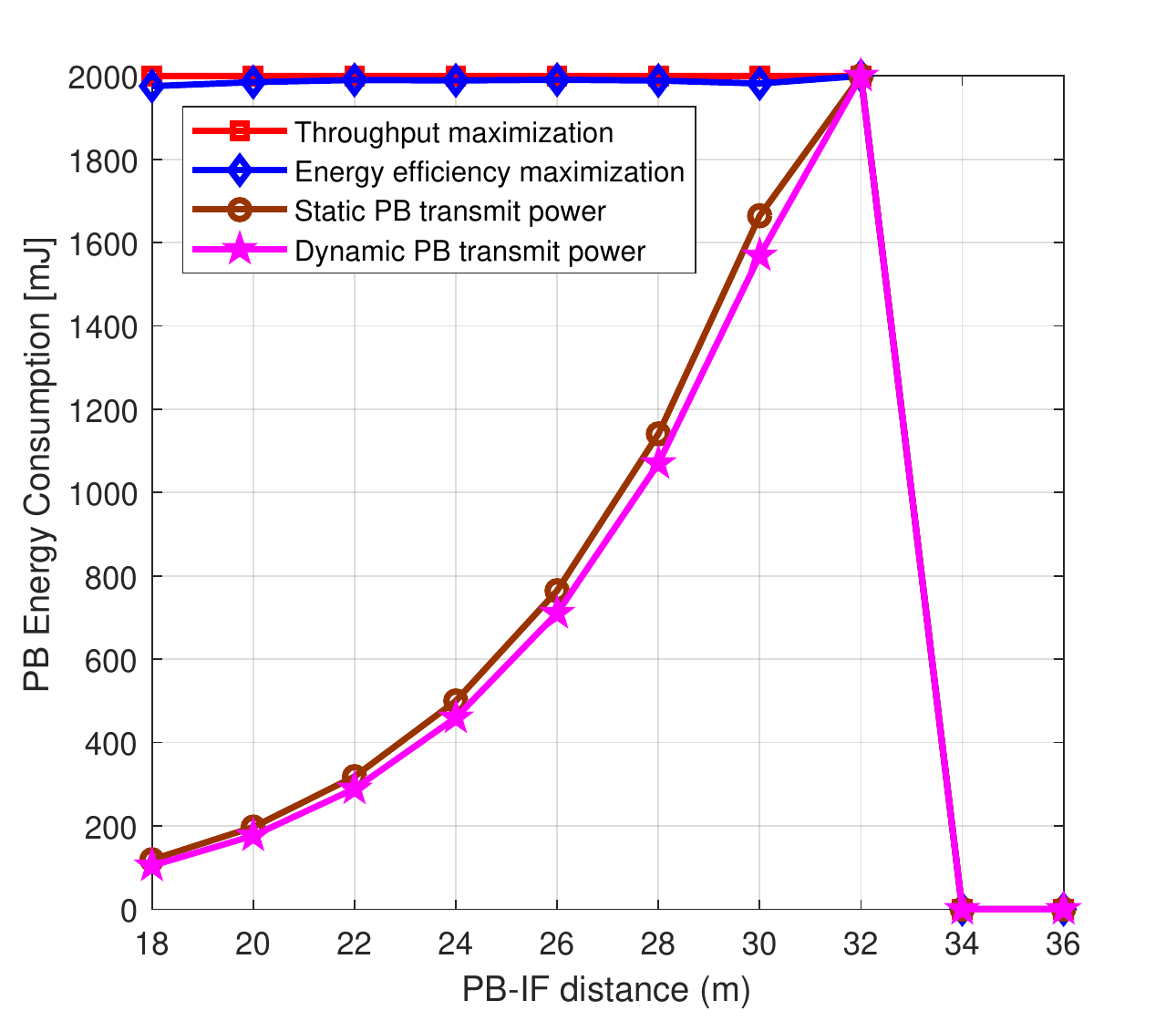}}
\caption{PB energy consumption versus the PB-IF distance $r$}
\end{figure}
Fig. 4 illustrates the PB energy consumption versus the PB-IF distance under the four different schemes. Both the throughput maximization and EE maximization schemes result in the maximum possible PB energy consumption in a time block for PB-IF distances up to 32 m, because they both require the use of the maximum PB transmit power and transmission time. For the proposed dynamic PB transmit power scheme and static PB transmit power scheme, the PB energy consumption increases with a longer PB-IF distance until it reaches 32 m, since the optimized PB transmit power and transmission time increase with the PB-IF distance. When the PB-IF distance goes beyond 32m, the PB energy consumption drops to 0, since none of the four schemes can find a feasible PB transmit power to compensate for the large path loss at long PB-IF distances, and each scheme returns a zero transmit power of the PB.

\begin{figure}[h]
\centerline{\includegraphics[width=3.5in]{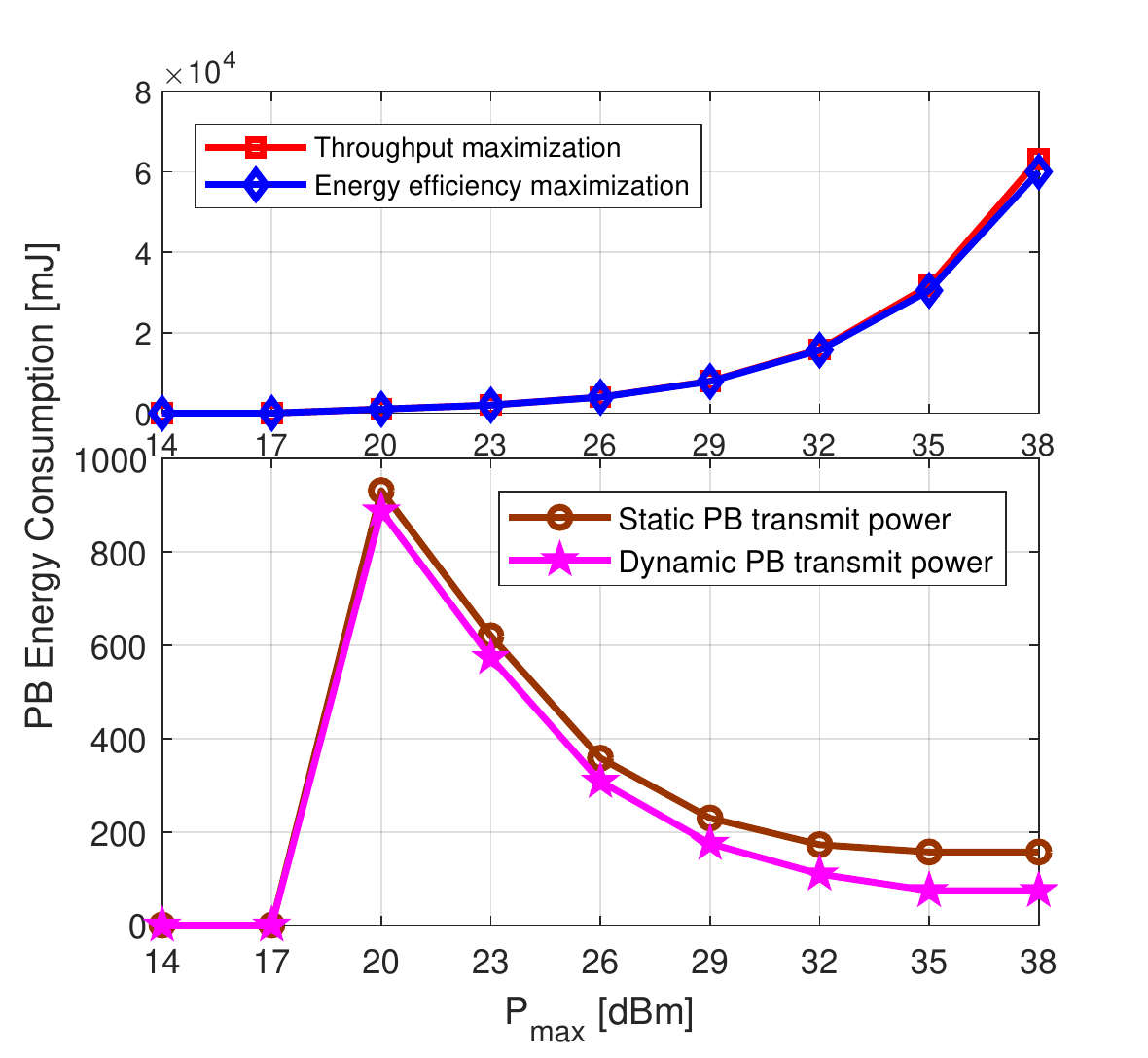}}
\caption{PB energy consumption versus $P_{\rm{max}}$ $(r=25m)$}
\end{figure}
Fig. 5 shows the PB energy consumption versus $P_{\rm{max}}$ under the four different schemes. We can see that when $P_{\rm{max}}$ is below 20 dBm, none of the four schemes can find a feasible value of PB transmit power and each sets it to 0. For $P_{max}>$ 20 dBm, the PB energy consumption under the throughput maximization and EE maximization schemes increases with $P_{max}$, because they both require the use of maximum possible PB transmit power and transmission time; while under the two proposed schemes, the PB energy consumption decreases with a higher $P_{\rm{max}}$ until it reaches 35 dBm. This is because a higher $P_{max}$ allows each IoT node to use a shorter transmission time $\tau_k$ to meet its throughput requirement, and thus a lower circuit energy consumption by each IoT node, which in turn leads to a lower PB transmit power $P_0$ during the pure EH phase $\tau_0$ in the dynamic PB transmit power scheme or a shorter duration $\tau_0$ of the pure EH phase in the static PB transmit power scheme. When $P_{\rm{max}}$ is above 35dBm, the PB energy consumption under the two proposed schemes becomes constant. This is because the optimal value of the PB transmit power under both schemes is well below $P_{\rm{max}}$ and the PB energy consumption is thus no longer affected by $P_{max}$. Furthermore, the gap between the PB energy consumption of the two proposed schemes becomes larger with $P_{\rm{max}}$ before it reaches 35dBm. This is because a higher value of $P_{\rm{max}}$ gives the dynamic PB transmit power scheme more flexibility to adjust the PB transmit power in each time slot, thereby further reducing the PB energy consumption as compared with the static PB transmit power scheme.


\section{Conclusion}
In this paper, we have proposed a dynamic PB transmit power scheme and a reduced complexity static PB transmit power scheme to minimize the PB energy consumption in a wireless powered BackCom network. By analyzing the KKT conditions, we find that the whole time block should be used up for EH and BackCom by all the IoT nodes in order to minimize the PB energy consumption in the dynamic PB transmit power scheme. However, minimizing the energy consumption of the PB in the static PB transmit power scheme does not require the whole time block to be used up. The simulation results show that the proposed algorithms converge very fast and that the dynamic PB transmit power scheme achieves the lowest PB energy consumption compared with the static PB transmit power and the benchmark schemes. The static PB transmit power scheme performs very close to the dynamic scheme but with a reduced implementation complexity. In our future work, we will consider multiple antennas on the PB and the IoT nodes for further improving the system performance.


\section*{Appendix A}
We can see from ${{\cal P}_{1.2}}$ that the objective function (9a) and the constraints  (9d) and (9e) are linear. This indicates that the problem ${{\cal P}_{1.2}}$ is convex if and only if the both constraints (9b) and (9c) are convex. The convexity of constraint (9b) depends on whether  $W{\tau _k}{\log _2}\left( {1 + \frac{{\xi {\lambda _k}{h_k}{g_k}}}{{{\tau _k}W{N_0}}}} \right) \ge R_k^{\min }$ is a joint concave function with respect to the optimization variables ${\lambda _k}$ and $\tau_k$. By using the convexity preserving of the perspective function, we know that the convexity of $W{\tau _k}{\log _2}\left( {1 + \frac{{\xi {\beta _k}{\theta _k}{h_k}{g_k}}}{{{\tau _k}W{N_0}}}} \right)$ is the same as $W{\log _2}\left( {1 + \frac{{\xi {\beta _k}{\theta _k}{h_k}{g_k}}}{{W{N_0}}}} \right)$, which is a concave function. Thus, the constraint (9b) is convex.

Next we prove that constraint (9c) is convex. For better understanding, we define the following two functions, i.e.,
\begin{align}
{f_1}\left( {{\theta _k},{\lambda _k},{\tau _k}} \right) = \left( {\frac{{\frac{{a\left( {{\theta _k} - {\lambda _k}} \right)}}{{{\tau _k}}}{h_k} + d}}{{\frac{{{\theta _k} - {\lambda _k}}}{{{\tau _k}}}{h_k} + v}} - \frac{d}{v}} \right){\tau _k},\tag{A.1}
\end{align}
\begin{align}
{f_2}\left( {{\theta _i},{\tau _i}} \right) = \left( {\frac{{a\frac{{{\theta _i}}}{{{\tau _i}}}{h_k} + d}}{{\frac{{{\theta _i}}}{{{\tau _i}}}{h_k} + v}} - \frac{d}{v}} \right){\tau _i}.\tag{A.2}
\end{align}
If both defined functions are joint concave functions with respect with the optimization variables, we can say that the constraint (9c) is concave. As above, it can be derived from the convexity preserving of the perspective function that the convexity of  ${f_1}\left( {{\theta _k},{\lambda _k},{\tau _k}} \right)$ and ${f_2}\left( {{\theta _i},{\tau _i}} \right)$ is determined by  ${f_3}\left( {{\theta _k},{\lambda _k}} \right) = \frac{{a\left( {{\theta _k} - {\lambda _k}} \right){h_k} + d}}{{\left( {{\theta _k} - {\lambda _k}} \right){h_k} + v}}- \frac{d}{v}$ and ${f_4}\left( {{\theta _i}} \right) = {\frac{{a{\theta _i}{h_k} + d}}{{{\theta _i}{h_k} + v}}}- \frac{d}{v}$, respectively. It can be observed that the convexity of  ${f_3}\left( {{\theta _k},{\lambda _k}} \right)$ and ${f_4}\left( {{\theta _i}} \right)$ is highly related with the adopted non-linear EH model ${f_5}\left( {x} \right) = {\frac{{ax + d}}{{x + v}}}- \frac{d}{v}$, where $x$ is the input power.

 As ${f_4}\left( {{\theta _i}} \right)$ is simpler than ${f_3}\left( {{\theta _k},{\lambda _k}} \right)$, in what follows, we first prove that ${f_4}\left( {{\theta _i}} \right)$ is a joint concave function and then put our attention on verifying the convexity of ${f_3}\left( {{\theta _k},{\lambda _k}} \right)$.
By comparing  ${f_4}\left( {{\theta _i}} \right)$ with ${f_5}\left( {x} \right)$, it is not hard to find that both of them have the same functional form and convexity. The  first- and second-order derivatives of ${f_5}\left( {x} \right)$ can be calculated as, respectively,
\begin{align}
\frac{{\partial {f_5}}}{{\partial x}}&{\rm{ = }}\frac{{av - d}}{{{{\left( {x + v} \right)}^2}}},\tag{A.3}\\
\frac{{{\partial ^2}{f_5}}}{{\partial {x^2}}}&{\rm{ = }}\frac{{2\left( {d - av} \right)}}{{{{\left( {x + v} \right)}^3}}}, \tag{A.4}
\end{align}
It has been well known that for a practical  energy harvester, the output power  increases with the input power first and then keeps  almost the same when the input power exceed the saturation threshold. Thus, $\frac{{\partial {f_5}}}{{\partial x}}> 0$, which means that ${av - d}> 0$. Based on this result, we know $\frac{{{\partial ^2}{f_5}}}{{\partial {x^2}}}<0$ if $x+v>0$. In the practical energy harvester, there exists a saturation threshold, i.e., $\mathop {\lim }\limits_{x \to \infty } {f_5}\left( x \right)=\frac{{av - d}}{v}>0$. Due to ${av - d}> 0$, we have $v>0$. Accordingly, $\frac{{{\partial ^2}{f_5}}}{{\partial {x^2}}}< 0$ and  ${f_4}\left( {{\theta _i}} \right)$ is a concave function with respect to ${\theta _i}$.

The Hessian matrix of ${f_3}\left( {{\theta _k},{\lambda _k}} \right)$ can be calculated as
\begin{align}\label{oo}
{\nabla ^2}{f_3}\left( {{\theta _k},{\lambda _k}} \right) = \left[ {\begin{array}{*{20}{c}}
{\frac{{2\left( {d - av} \right)h_k^2}}{{{{\left[ {\left( {{\theta _k} - {\lambda _k}} \right){h_k} + v} \right]}^3}}}}&{\frac{{2\left( {av - d} \right)h_k^2}}{{{{\left[ {\left( {{\theta _k} - {\lambda _k}} \right){h_k} + v} \right]}^3}}}}\\
{\frac{{2\left( {av - d} \right)h_k^2}}{{{{\left[ {\left( {{\theta _k} - {\lambda _k}} \right){h_k} + v} \right]}^3}}}}&{\frac{{2\left( {d - av} \right)h_k^2}}{{{{\left[ {\left( {{\theta _k} - {\lambda _k}} \right){h_k} + v} \right]}^3}}}}
\end{array}} \right]\tag{A.5}
\end{align}
It is not hard to verify that the Hessian matrix in \eqref{oo} is  negative semidefinite and thus ${f_3}\left( {{\theta _k},{\lambda _k}} \right)$ is a joint concave function.

Based on the above analysis, we confirm that constraint (9c) is convex, and the proof is complete.

\section*{Appendix B}
According to the KKT conditions, $\frac{{\partial {\cal L}}}{{\partial {\theta _0}}}=1 - \sum\limits_{k = 1}^K {\frac{{{\alpha _k}\left( {av - d} \right){h_k}}}{{{{\left( {\frac{{{\theta _0}}}{{{\tau _0}}}{{h_k}} + v} \right)}^2}}}}  + {\omega _0}=0$ should be satisfied for minimizing the energy consumption of the PB. As proven in Appendix A, ${av - d}>0$ holds. Combining $\frac{{\partial {\cal L}}}{{\partial {\theta _0}}}=0$,  ${av - d}>0$, and ${\omega _0}\ge0$, it is not hard to infer that  there is at least one IoT node $\hat{k}$ satisfying $\alpha_{\hat{k}}>0$. For the $\hat{k}$-th IoT node, $\frac{{\partial {\cal L}}}{{\partial {\lambda _{\hat{k}}}}} = 0$ should be satisfied. Due to ${\kappa _{\hat{k}}}\ge 0$, we have ${\varepsilon _{\hat{k}}>0}$. Using the
complementary slackness conditions ${\alpha _{\hat{k}}}\bigg[ {p_{c,{\hat{k}}}}{\tau _{\hat{k}}} - \left( {\frac{{\frac{{a\left( {{\theta _{\hat{k}}} - {\lambda _{\hat{k}}}} \right)}}{{{\tau _{\hat{k}}}}}{h_{\hat{k}}} + d}}{{\frac{{{\theta _{\hat{k}}} - {\lambda _{\hat{k}}}}}{{{\tau _{\hat{k}}}}}{h_{\hat{k}}} + v}} - \frac{d}{v}} \right){\tau _{\hat{k}}} - \sum\limits_{i = 0,i \ne {\hat{k}}}^K {\left( {\frac{{a\frac{{{\theta _i}}}{{{\tau _i}}}{h_{\hat{k}}} + d}}{{\frac{{{\theta _i}}}{{{\tau _i}}}{h_{\hat{k}}} + v}} - \frac{d}{v}} \right){\tau _i}}  \bigg]=0$ and ${\varepsilon _{\hat{k}}}\left[ {R_{\hat{k}}^{\min } - W{\tau _{\hat{k}}}{{\log }_2}\left( {1 + \frac{{\xi {\lambda _{\hat{k}}}{h_{\hat{k}}}{g_{\hat{k}}}}}{{{\tau _{\hat{k}}}W{N_0}}}} \right)} \right]{\rm{ = }}0$, the equality should be satisfied in both constraints ${\rm{C}}1''$ and ${\rm{C}}2''$ for optimality solving $\mathcal{P}_{1.2}$, which results in Insight 1. The proof is complete.

\section*{Appendix C}
We prove this Insight from the following two steps.

Step 1:  As shown in Appendix B, there is at least one IoT node $\hat{k}$ ($\hat{k}\in \{1,2,...,K\}$) satisfying $\alpha_{\hat{k}}>0$. Since $\frac{{\partial {\cal L}}}{{\partial {\tau _0}}} =   - \sum\limits_{k = 1}^K {\alpha _k}\left( {a - \frac{d}{v} + \frac{{\left( {d - av} \right)\left( {\frac{{2{\theta _0}}}{{{\tau _0}}}{h_k} + v} \right)}}{{{{\left( {\frac{{{\theta _0}}}{{{\tau _0}}}{h_k} + v} \right)}^2}}}} \right)
 + \vartheta- {\omega _0}{P_{\max }}=0$,   $\vartheta>0$ holds if the following inequality \eqref{B1} is satisfied.
\begin{align}\label{B1}
 a - \frac{d}{v} + \frac{{\left( {d - av} \right)\left( {\frac{{2{\theta _0}}}{{{\tau _0}}}{h_k} + v} \right)}}{{{{\left( {\frac{{{\theta _0}}}{{{\tau _0}}}{h_k} + v} \right)}^2}}} > 0.\tag{C.1}
 \end{align}
 To prove \eqref{B1}, we construct the following function, i.e.,
 \begin{align}
f_6\left( {{h_k}} \right) = a - \frac{d}{v} + \frac{{\left( {d - av} \right)\left( {\frac{{2{\theta _0}}}{{{\tau _0}}}{h_k} + v} \right)}}{{{{\left( {\frac{{{\theta _0}}}{{{\tau _0}}}{h_k} + v} \right)}^2}}}.\tag{C.2}
 \end{align}
The first-order derivative of $f_6\left( {{h_k}} \right)$ is calculated as
\begin{align}
f'_6\left( {{h_k}} \right) =\left( {av - d} \right)\frac{{2{{\left( {\frac{{{\theta _0}}}{{{\tau _0}}}} \right)}^2}{h_k}}}{{{{\left( {\frac{{{\theta _0}}}{{{\tau _0}}}{h_k} + v} \right)}^3}}}.\tag{C.3}
\end{align}
 Since ${av-d}>0$ and $v>0$, it is not hard to verify that $f'_6\left( {{h_k}} \right)>0 $ when  ${h_k}\ge 0$. In the practical communication scenario, we have $h_k>0$, which results in  $f_6\left( {{h_k}} \right)> f_6\left( {0} \right)=0$, i.e., the inequality \eqref{B1} holds.

 Step 2: Combining the complementary slackness condition $\vartheta \left[ {\sum\limits_{i = 0}^K {{\tau _i}}  - T} \right]=0$ and the result $\vartheta>0$, we have $ {\sum\limits_{i = 0}^K {{\tau _i}}  = T} $, yielding Insight 2.

The proof is complete.

\section*{Appendix D}
Let us define a function given by $f_1=\left( {\frac{{a\left( {1 - {Y_k}} \right){P}{h_k} + d}}{{\left( {1 - {Y_k}} \right){P}{h_k} + v}} - \frac{d}{v}} \right){\tau _k}$, where $f_1$ is convex with respect to $Y_k$. Since the first-order Taylor expansion of a convex function is a global under-estimator of its function values. For any given $Y_k^j$, we have
\begin{align}
f_1 &\ge \left( {\frac{{a\left( {1 - {Y_k^j}} \right){P}{h_k} + d}}{{\left( {1 - {Y_k^j}} \right){P}{h_k} + v}} - \frac{d}{v}} \right){\tau _k}\notag\\
&+\frac{Ph_kd-aPh_kv}{(Ph_k-Y_k^jPh_k+v)^2}(L_k-Y_k^j\tau_k),\tag{D.1}
 \end{align}
where the equalities in (D.1) hold when $\frac{L_k}{\tau_k}=Y_k^j$.

The proof is complete.

\section*{Appendix E}
We give the partial Lagrangian function for ${\cal P}_{2.1}$, which is a convex problem, as
\begin{equation}\tag{E.1}
\begin{split}
\mathcal{L^{'}} =& \sum\limits_{i = 0}^{K} {P\tau_i} +\sum\limits_{k = 1}^K {{\alpha _k^{'}}\left( {{p_{c,k}}{\tau _k}-E_k^{\rm{total{'}}}} \right)}+\vartheta^{'} \left( {\sum\limits_{i = 0}^K {{\tau _i}}  - T} \right)\\
&+\sum\limits_{k = 1}^K {{\varepsilon _k^{'}}\left[ {R_k^{\min } - W{\tau _k}{{\log }_2}\left( {1 + \frac{{\xi {L_k}{h_k}{g_k}}}{{{\tau _k}W{N_0}}}} \right)} \right]}\\
&{\rm{ + }}\sum\limits_{k = 1}^K {{\kappa _k^{'}}\left( {{L _k} - {\tau _k}} \right)},
\end{split}
\end{equation}
where $\alpha _k^{'}, \vartheta^{'}, \varepsilon _k^{'}$ and $\kappa _k^{'}$ are the Lagrange multipliers associated with the constraints in ${\cal P}_{2.1}$.

Taking the partial derivative of $\mathcal{L^{'}}$ with respect to $\tau_{i}, i\in \left[0, 1, 2, \cdots, K\right]$, there always exists two terms, i.e, $P+\vartheta^{'}$. Since PB transmit power $P>0, \vartheta^{'}\ge 0$ and based on KKT conditions, $\vartheta^{'}$ could be 0 while satisfying $\frac{{\partial {\cal L^{'}}}}{{\partial {\tau _i}}}=0$. Therefore, ${\sum\limits_{i = 0}^K {{\tau _i}}}$ could not be equal to $T$ for meeting the complementary slackness condition, that is to say, the available time could not be used up for minimizing the energy consumption of the PB.

The proof is complete.


\ifCLASSOPTIONcaptionsoff
  \newpage
\fi
\bibliographystyle{IEEEtran}
\bibliography{refa}

\end{document}